\documentclass[a4paper,11pt]{article}
\usepackage{pos}
\usepackage{dirtytalk}
\usepackage{caption}
\usepackage{subcaption}
\usepackage{wrapfig}
\usepackage{tikz}

\newcommand{\Z}{{\mathbb{Z}}}
\newcommand{\R}{{\mathbb{R}}}

\newcommand{\p}{\partial}

\newcommand{\ket}[1]{\lvert #1 \rangle}

\newcommand{\U}{\mathrm{U}}

\newcommand{\pqty}[1]{\left( #1 \right)}
\newcommand{\bqty}[1]{\left[ #1 \right]}
\newcommand{\abs}[1]{\left\lvert #1\right\rvert}

\newcommand{\expval}[1]{\langle #1 \rangle}

\newcommand {\dv}[3][ ]{
  \ifx #1 { }
    \frac{d #2}{d #3}
  \else
    \frac{d^{#1} #2}{d #3^{#1}}
  \fi
}
\newcommand {\pdv}[3][ ]{
  \ifx #1 { }
    \frac{\partial #2}{\partial #3}
  \else
    \frac{\partial^{#1} #2}{\partial #3^{#1}}
  \fi
}
\newcommand {\fdv}[3][ ]{
  \ifx #1 { }
    \frac{\delta #2}{\delta #3}
  \else
    \frac{\delta^{#1} #2}{\delta #3^{#1}}
  \fi
}

\title{A New Type of Lattice Gauge Theory through Self-adjoint Extensions
}

\author[a,b]{A. Banerjee}
\author[a,b]{D. Banerjee}
\author[c]{G. Kanwar}
\author*[c]{A. Mariani}
\author[c]{T. Rindlisbacher}
\author[c]{U.J. Wiese}

\affiliation[a]{Theory Division, Saha Institute of Nuclear Physics, 1/AF\\
Bidhan Nagar,\\
Kolkata 700064, India}

\affiliation[b]{Homi Bhabha National Institute, Training School Complex,\\
Anushaktinagar,\\
Mumbai 400094,India}

\affiliation[c]{Albert Einstein Center for Fundamental Physics,\\ Institute for Theoretical Physics, Bern University\\
Sidlerstrasse 5, CH-3012 Bern, Switzerland}

\emailAdd{aditya.banerjee@saha.ac.in}
\emailAdd{debasish.banerjee@saha.ac.in}
\emailAdd{kanwar@itp.unibe.ch}
\emailAdd{mariani@itp.unibe.ch}
\emailAdd{trindlis@itp.unibe.ch}
\emailAdd{wiese@itp.unibe.ch}

\abstract{A generalization of Wilsonian lattice gauge theory may be obtained by considering the possible self-adjoint extensions of the electric field operator in the Hamiltonian formalism. In the special case of 3D $\U(1)$ gauge theory these are parametrised by a phase $\theta$, and the ordinary Wilson theory is recovered for $\theta=0$. We consider the case $\theta=\pi$, which, upon dualization, turns into a theory of staggered integer and half-integer height variables. We investigate order parameters for the breaking of the relevant symmetries, and thus study the phase diagram of the theory, which shows evidence of a broken $\Z_2$ symmetry in the continuum limit, in contrast to the ordinary theory.}

\FullConference{%
  The 39th International Symposium on Lattice Field Theory, LATTICE2022\\ 8th-13th August, 2022\\
  University of Bonn, Bonn, Germany
}

\begin{document}
\maketitle

\section{Introduction}

The action of a quantum field theory is determined by the symmetries of the physical system it is meant to model. In the Wilsonian framework of renormalization, for example, one includes in principle all possible terms consistent with the symmetries of the theory. In the context of gauge theories, the most important symmetries are Poincaré invariance and the gauge symmetry. On the lattice the gauge symmetry is typically implemented exactly, while Poincaré invariance is broken to the subgroup of lattice translations and space-time hypercubic rotations. One then expects to recover the full Poincaré group in the continuum limit. 

In the typical framework of gauge theories, one may consider different choices of action with the gauge symmetry and lattice Poincaré symmetry, all of which converge to the \textit{same} continuum limit. In the present work we consider an example of a new type of gauge theory - in particular, a $\U(1)$ gauge theory in three dimensions - which, despite preserving both the lattice gauge symmetry and the lattice Poincaré invariance, has a different continuum limit than the standard Wilson-type $\U(1)$ gauge theory. In fact, by formulating the standard $\U(1)$ gauge theory in the Hamiltonian formalism, we see that it is natural to include an extra parameter in the theory, in this case an angle $\theta$. This extension, which in a mathematical sense corresponds to a \textit{self-adjoint extension} of the electric field operator in the Hamiltonian formalism, would be far from obvious from a Euclidean action perspective. In this sense, our work expands the current framework of lattice gauge theories. 

In these proceedings, we first formulate the new type of $\U(1)$ gauge theory and explain how the $\theta$ parameter is introduced. In the $\theta=0$ case, it reduces to the standard $\U(1)$ gauge theory, for which we briefly review the main results in the three-dimensional case. We specialize to three dimensions for simplicity and then consider the case $\theta=\pi$, which is the only value of $\theta$ which preserves the symmetries of the $\theta=0$ theory; in particular, for $\theta \not\in \{0, \pi\}$, charge conjugation is explicitly broken. Similarly to the standard $\U(1)$ gauge theory, the $\theta=\pi$ theory may be written in terms of a dual height model, which we simulate numerically. The dual model is free of the sign problem that appears in the path integral formulation of the $\theta=\pi$ theory. We investigate appropriate order parameters for the breaking of the relevant symmetries, and we find evidence that the $\theta=\pi$ theory has a broken $\Z_2$ symmetry down to the continuum limit, a feature which is absent from the standard $\U(1)$ theory.

\section{The standard compact Abelian gauge theory in 3D}\label{sec:usual u(1)}

In this section we briefly review some results, as well as the Hamiltonian formulation, for the standard compact $\U(1)$ gauge theory in three dimensions.

\subsection{Action formulation}

The standard compact Wilson-type $\U(1)$ gauge theory in 3D is well-understood both analytically and numerically \cite{GopfMack}. In the Villain formulation, its partition function is given by
\begin{equation}
    \label{eq:usual u(1) partition function}
    Z = \pqty{\prod_{l \in \mathrm{links}} \int_{-\pi}^\pi d\theta_l} \pqty{\prod_{p \in \mathrm{plaq}} \sum_{n_p = -\infty}^{+\infty}} \exp{\bqty{ -\frac{1}{2e^2} \sum_p ((d\theta)_p - 2\pi n_p)^2}} \ ,
\end{equation}
where $e^2$ is a dimensionless coupling, $l$ are lattice links and $p$ are plaquettes. Labelling the ordered links in the plaquette $p$ from $1$ to $4$, we have $(d\theta)_p = \theta_1 + \theta_2 - \theta_3 - \theta_4$. Many features of the theory can be understood analytically by dualization. The partition function eq.\eqref{eq:usual u(1) partition function} can be rewritten in terms of new field variables $h_x \in \Z$ which live on the sites $x$ of the dual lattice. The end result is that, up to some constant prefactors,
\begin{equation}
    \label{eq:unstaggered height model}
    Z = \pqty{\prod_{x \in \mathrm{sites}} \sum_{h_x=-\infty}^{+\infty}}\exp{\bqty{-\frac{e^2}{2} \sum_{\expval{xy}} (h_x-h_y)^2}} \ ,
\end{equation}
where $h_x \in \Z$ are integer-valued scalar fields on the dual lattice, and $\expval{xy}$ denotes nearest neighbours. It is precisely the integer nature of the fields which makes the theory non-trivial. If $h_x \in \R$, this would simply be the action of a massless free real scalar field.

Through the analysis of the height model eq.\eqref{eq:unstaggered height model}, it has been shown analytically that 3D $\U(1)$ gauge theory is confining at all couplings, with a mass gap $m$ which scales as \cite{GopfMack}
\begin{equation}
    \label{eq:mass gap}
    m^2 = \frac{8\pi^2}{e^2} \exp{\bqty{-2\pi^2 c/e^2}} \ ,
\end{equation}
in lattice units, where $c \approx 0.2527$, while the string tension scales as
\begin{equation}
    \sigma = \frac{\widetilde{c}}{4\pi^2} m e^2 
\end{equation}
for some constant $\widetilde{c}$, again in lattice units. These results have found numerical confirmation \cite{TepAthen, CasPan}. The continuum limit is achieved as $e^2 \to 0$. Perhaps surprisingly, the resulting continuum theory is a free scalar field of mass $m$ \cite{GopfMack}. 

\subsection{Hamiltonian formulation}

In the Hamiltonian formulation of $\U(1)$ gauge theory, time is continuous while space is discretized into a square lattice \cite{KogSuss}. The temporal gauge $A_0=0$ is chosen. Classically, one assigns a $\U(1)$ variable to each lattice link. Thus, quantising, the Hilbert space on each link is given by the set of square-integrable functions on $\U(1)$ and the overall Hilbert space is therefore
\begin{equation}
    \mathcal{H} = \bigotimes_{l \in \mathrm{links}} L^2(\U(1))_l \ .
\end{equation}
The Hamiltonian of the theory can be obtained via the transfer matrix and is given by
\begin{equation}
    H = \frac{e^2}{2} \sum_{l \in \mathrm{links}} E_l^2 + \frac{1}{2e^2}\sum_{p \in \mathrm{plaq}} B^2(U_p) \ ,
\end{equation}
where $U_p$ is the usual plaquette variable and the electric field operator $E_l$ is given on each link by $E_l = -i\pdv{}{\varphi_l}$, where $U_l = \exp{(i\varphi_l)}$ is the $\U(1)$ link variable. Thus one may interpret $\varphi_l$ as an angular position operator on the circle $\U(1)$, with $E_l$ its canonically conjugate angular momentum. The magnetic term $B^2$ may be chosen in different equivalent ways, similarly to the freedom available in the choice of action. For example, $B^2$ could be either the Wilson or Villain term for the standard $\U(1)$ gauge theory. There is an additional constraint on the Hilbert space, in that only those states which satisfy the Gauss' law constraint
\begin{equation}
    \label{eq:gauss law}
    G_x \ket{\psi} = 0 \ , \quad\quad\quad\quad G_x = \sum_{i} \pqty{E_{x,i}-E_{x-i,i}} 
\end{equation}
are to be considered as physical states.

\section{A new type of Abelian gauge theory}

In the previous section we have seen that, in the Hamiltonian formulation, on each lattice link the Hilbert space is formed by the square-integrable functions on $\U(1)$. Each such wavefunction $\psi \in L^2(\U(1))$ therefore satisfies
\begin{equation}
    \label{eq:periodic bcs}
    \psi(2\pi)=\psi(0), \quad \quad \quad \quad \int_0^{2\pi} d\varphi \abs{\psi(\varphi)}^2 < \infty \ .
\end{equation}
In other words, on each link the wavefunctions are required to come back to themselves after going around $2\pi$ in gauge field space. This, however, is not the most general choice that is consistent with the symmetries of the theory. One may, in fact, require that the wavefunction only come back to itself up to a twist,
\begin{equation}
    \label{eq:twisted periodic bcs}
    \psi(2\pi)=e^{i\theta}\psi(0) \ .
\end{equation}
This choice is consistent with the gauge symmetry, so that the resulting theory still has a $\U(1)$ gauge symmetry. In particular, the choice eq.\eqref{eq:twisted periodic bcs} is still compatible with the Gauss law eq.\eqref{eq:gauss law}. Another way of understanding the choice eq.\eqref{eq:twisted periodic bcs} is through the mathematical concept of \textit{self-adjoint extensions} \cite{ReedSimon, Gieres}. In the standard Wilson-type gauge theory, the electric field $E=-i\partial_\varphi$ is an operator on the single-link Hilbert space. In particular, in order to make sense of the Hamiltonian and the Gauss' law constraint, the electric field must be \textit{self-adjoint}, so that it has a complete orthonormal basis of eigenfunctions and real eigenvalues. The choice eq.\eqref{eq:twisted periodic bcs} is in fact the most general choice consistent with the requirement that $E=-i\partial_\varphi$ be self-adjoint.

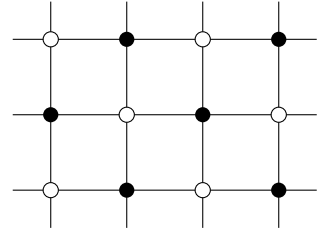
\begin{wrapfigure}{r}{0.4\textwidth}
    \centering
    \begin{tikzpicture}
        \draw[step=1.0,black] (0.5,0.5) grid (4.5,3.5);
        
        \foreach \p in {(2,1),(2,3),(1,2),(3,2), (4,1), (4,3)}
            \fill \p circle(.10);
        
        \foreach \p in {(1,1),(1,3),(2,2),(3,1), (3,3), (4,2)}
        {
            \fill[white] \p circle(.10);
            \draw \p circle(.10);
        }
        
    \end{tikzpicture}
    \caption{A two dimensional slice of the lattice, depicting the staggered integer (black dots) and half-integer (white dots) height variables.}
    \label{fig:staggered lattice}
\end{wrapfigure}

Up until now, we have modified only the domain of definition of the electric field operator, i.e. the Hilbert space of the theory. In fact, requiring that the wavefunctions satisfy eq.\eqref{eq:twisted periodic bcs} in this case is equivalent, by a basis change, to having wavefunctions which satisfy eq.\eqref{eq:periodic bcs} with a modified electric field operator $E_l \to E_l +\tfrac{\theta}{2\pi}$. Starting from the Villain form of the magnetic term $B^2$ of the Hamiltonian, we then modify it, similarly to the electric field, in such a way to preserve both the structure of small fluctuations and the overall Euclidean lattice cubic symmetry in the partition function. Since we aim to understand the resulting theory numerically, we obtain the partition function from the Hamiltonian. We then dualize the partition function, as we've seen in Section \ref{sec:usual u(1)} for the standard $\U(1)$. We note that for $\theta \neq 0$, the original partition function has a sign problem which is absent from the dualized partition function expressed in terms of the dual variables.

Among the possible values of $\theta$, only the $\theta = \pi$ theory preserves all the symmetries of the $\theta = 0$ theory; in particular, for $\theta \not\in \{ 0, \pi \}$ charge conjugation is explicitly broken. We therefore focus on the $\theta=\pi$ case, for which the dualized partition function is that of a height model,
\begin{equation}
    \label{eq:staggered height model}
    Z = \pqty{\prod_{x \mathrm{even}} \sum_{h_x \in \Z}} \pqty{\prod_{x \mathrm{odd}} \sum_{h_x \in \Z + 1/2}}\exp{\bqty{-\frac{e^2}{2} \sum_{\expval{xy}} (h_x-h_y)^2}} \ ,
\end{equation}
where each lattice site at position $\vec x = (x_0, x_1, x_2)$ in lattice units is said to be even or odd according to the parity of $x_0+x_1+x_2$. The height variables $h_x$ are integer on even lattice sites and half-integer on odd lattice sites. The staggering is isotropic in space-time; each integer variable is surrounded by half-integer nearest neighbours in all directions, and viceversa. This should be contrasted with the case of quantum link models, where a similar dualization can be performed but the staggering only occurs in the space directions \cite{QuantumLink}. An example of a staggered lattice in two dimensions is shown in figure \ref{fig:staggered lattice}, however in our case we work in three dimensions and the staggering is in all directions. The partition function eq.\eqref{eq:staggered height model} should be contrasted with the dual partition function for the standard $\U(1)$ theory, eq.\eqref{eq:unstaggered height model}; the two are identical except for the nature of the height variables, which for the latter are all integer-valued. 

\section{Symmetries and order parameters}

The theory described by the partition function eq.\eqref{eq:staggered height model}, i.e. the dualized $\theta=\pi$ 3D $\U(1)$ gauge theory, has several symmetries:
\begin{enumerate}
    \item \textit{Global $\Z$-invariance}: $h_x \to h_x + c$ where $c$ is any constant integer. This preserves both the action and the integer and half-integer nature of the height variables.
    \item \textit{$\Z_2$ charge conjugation $C$}: $h_x \to -h_x$. Again this preserves both the action and the nature of the height variables.
    \item \textit{$\Z_2$ shift symmetry $S$}: $h_x \to h_{x+\mu}+\frac12$ where $\mu$ is any direction. Since shifting by one lattice spacing in any direction replaces integer height variables with half-integers and vice-versa, we add a half-integer shift in order to restore their nature. 
    \item \textit{Translations by an even number of lattice spacings}. These preserve the nature of the height variables and therefore do not require any further staggering.
    \item \textit{Euclidean cubic symmetry}. The action is isotropic with respect to lattice rotations and reflections in any of the three directions.
\end{enumerate}
The global $\Z$-invariance should be thought of as a redundancy in our description of the system, whereby we are free to set the overall height of the system \cite{GopfMack}. We will therefore require that all observables be invariant under the global $\Z$-symmetry.  

We now construct order parameters for the possible breaking of the $S$ and $C$ symmetries. We consider first the observable
\begin{equation}
    O_{CS} = \sum_{x} (-1)^x h_x = \sum_{x\,\mathrm{even}} h_x - \sum_{x\,\mathrm{odd}} h_x \ ,
\end{equation}
which is sensitive to both $S$ and $C$, under either of which it changes sign. We note that $O_{CS}$ is indeed invariant under the global $\Z$-symmetry, and moreover, it is a sum of local observables.

An observable which is sensitive to $S$ but not $C$ is more difficult to construct, since it must also be $\Z$-invariant and the sum of local terms. We chose the observable $O_S$ to be
\begin{equation}
    O_S = \sum_{c \in \mathrm{cubes}} \sum_{x \in c} (-1)^x (h_x - \bar{h}_c)^2 \ ,
\end{equation}
where the sum is first over all unit cubes in the lattice and then over all sites $x$ belonging to cube $c$ with the relevant parity. The cube average $\bar{h}_c$ is defined as
\begin{equation}
    \bar{h}_c = \frac{1}{8} \sum_{x \in c} h_x \ .
\end{equation}
The observable $O_S$ is invariant under charge conjugation $C$ but changes sign under the shift symmetry $S$. We note that a theory whose phase diagram was characterized by order parameters for the breaking of charge conujugation and one-site shift symmetry was already considered in the context of quantum link models \cite{QuantumLink}.

\section{Numerical simulation}

\begin{figure}
    \centering
    \includegraphics{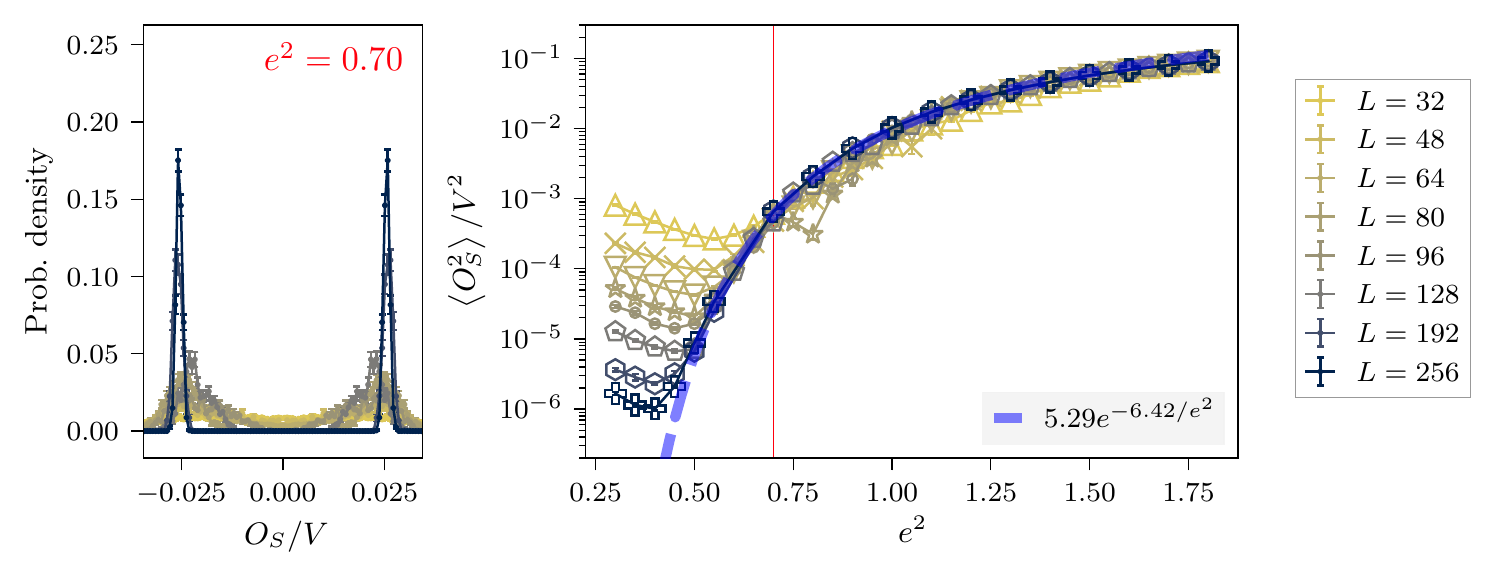}
    \caption{Left: Histogram of the operator $O_S$ normalized by the volume $V$ at $e^2=0.70$. Right: Susceptibility of $O_S$ normalized by $V^2$ as a function of $e^2$ for several volumes, with a fit of the large-volume data to the form $A \exp{(-B / e^2)}$. The double-peaked structure at large volume (left) and scaling as $V^2$ (right) indicates spontaneous breaking of the $S$ symmetry.}
    \label{fig:OS observable data}
\end{figure}

\begin{figure}
    \centering
    \includegraphics{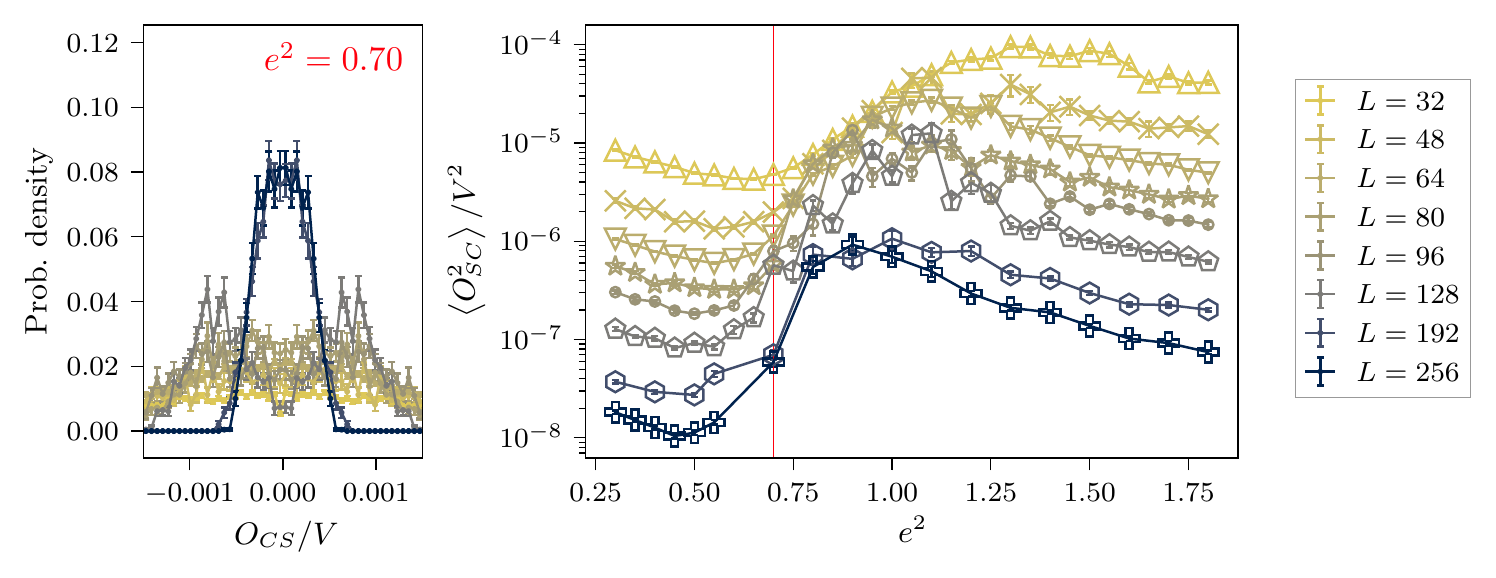}
    \caption{Left: Histogram of the operator $O_{CS}$ normalized by the volume at $e^2=0.70$. Right: Susceptibility of $O_{CS}$ normalized by $V^2$ as a function of $e^2$ for several volumes. A single histogram peak at large volumes (left) and scaling smaller than $V^2$ (right) indicates an unbroken $CS$ symmetry.}
    \label{fig:OCS observable data}
\end{figure}

We simulated the staggered height model numerically using a multi-cluster algorithm \cite{Evertz}. We investigated a range of couplings from $e^2 = 0.3$ up to $e^2 = 2.0$ on $L^3$ lattices from $L=32$ up to $L=256$. We computed the observables $O_S$ and $O_{CS}$ and show the relevant data in figures \ref{fig:OS observable data} and \ref{fig:OCS observable data}. We expect the continuum limit to emerge as $e^2 \to 0$, as in the case of the standard $\U(1)$ gauge theory.

Focusing on figure \ref{fig:OS observable data}, we see that the susceptibility of $O_S$, appropriately normalized by the volume squared, is essentially volume-independent for a wide range of coupling $e^2$, indicating that the $S$ symmetry is spontaneously broken. This is confirmed by the histogram, which shows two clearly defined peaks which become sharper with increasing volume. As the coupling becomes smaller, finite volume effects cause the $S$ symmetry to be restored. However, we see that this happens for decreasing values of $e^2$ as the volume is increased, which indicates that this is likely a finite-volume effect. Moreover, the function $f(e^2)=A \exp{(-B/e^2)}$, inspired by eq.\eqref{eq:mass gap}, provides an excellent fit to the data and would imply that the symmetry is broken for all couplings. We interpret the data to mean that the $S$ symmetry (shift by one lattice spacing) is broken for all values of the coupling $e^2$. It is important to note that translations by an even number of lattice spacings remain unbroken, so that we still expect a translational invariant theory in the continuum limit.

The data for the $O_{CS}$ observable, on the other hand, is shown in figure \ref{fig:OCS observable data}. In this case the susceptibility normalized by $V^2$ decreases with the volume, which is indicative that at least one of $S$ or $C$ is unbroken. The decrease with the volume is less apparent in the central region of the right-hand side of figure \ref{fig:OCS observable data}, but the histogram in figure \ref{fig:OCS observable data}, taken at $e^2=0.70$ (in this region) shows that, while two peaks seem to be forming for small volumes, they merge in the center around zero for bigger volumes. We therefore interpret the data to mean that the $C$ symmetry remains unbroken for all couplings.

Overall, we therefore see evidence that the continuum limit of the unstaggered model, obtained as $e^2 \to 0$, has a $\Z_2$ broken symmetry which remains unbroken in the standard Wilson-type $\U(1)$ gauge theory. It is conceivable that the broken shift symmetry may manifest itself in terms of the internal degrees of freedom of the continuum theory. It would be interesting to obtain an effective description of the $\theta=\pi$ theory in the continuum and compare with the $\theta=0$ case, which, as we have seen in Section \ref{sec:usual u(1)}, is described in the continuum by a massive free scalar.

\section{Conclusions}

We have shown that $\U(1)$ gauge theory may be extended by the addition of an extra non-perturbative parameter $\theta$, in a way which preserves both the gauge symmetry and the Euclidean hypercubic symmetry. We considered the theory in three dimensions for $\theta=\pi$ and dualized it to obtain a height model free of the sign problem which affects the original theory. We find evidence that the $\theta=\pi$ theory has a broken $\Z_2$ symmetry which is absent from the standard $\U(1)$ gauge theory, and therefore describes a different continuum limit.

We are currently performing additional numerical work to compute other quantities for the staggered height model, such as the mass gap and the string tension. It would also be worthwhile to understand the universality class corresponding to the continuum limit of the $\theta=\pi$ theory, and understand the nature of the effective theory that describes it.

Other interesting extensions of the present work involve studying the theory in different dimensions (for example in 4D) or at values of $\theta \not\in \{0, \pi\}$ and, perhaps more importantly, understanding the theoretical framework and performing numerical simulations in the non-Abelian case.

\acknowledgments

GK, AM and UJW acknowledge funding from the Schweizerischer Nationalfonds (grant agreement number 200020\_200424).


\begin{thebibliography}{99}

\bibitem{GopfMack}
M. G\"opfert, G. Mack,
\emph{Proof of confinement of static quarks in 3-dimensional U(1) lattice gauge theory for all values of the coupling constant},
\emph{Comm. Math. Phys.} {\bf 82} (1982) 4

\bibitem{TepAthen}
A. Athenodorou, M. Teper,
\emph{On the spectrum and string tension of U(1) lattice gauge theory in 2 + 1 dimensions},
\emph{JHEP} {\bf 2019} (2019) 63
[{\tt arXiv:1811.06280}].

\bibitem{CasPan}
M. Caselle, M. Panero, R. Pellegrini, and D. Vadacchino,
\emph{A different kind of string},
\emph{JHEP} {\bf 2015} (2015) 105
[{\tt 	arXiv:1406.5127}].

\bibitem{KogSuss}
J. Kogut, L. Susskind,
\emph{Hamiltonian formulation of Wilson's lattice gauge theories},
\emph{Phys. Rev. D} {\bf 11} (1975) 395.

\bibitem{ReedSimon}
M. Reed and B. Simon,
\emph{II: Fourier Analysis, Self-Adjointness: Volume 2. Methods of Modern Mathematical Physics},
San Diego, CA: Academic Press, Nov. 1975.

\bibitem{Gieres}
F. Gieres,
\emph{Mathematical surprises and Dirac's formalism in quantum mechanics},
\emph{Rep.Prog.Phys.} {\bf 63} (2000) 1893
[{\tt arXiv:quant-ph/9907069}].

\bibitem{QuantumLink}
D. Banerjee, F.-J. Jiang, P. Widmer, and U.-J. Wiese,
\emph{The (2 + 1)-d U(1) quantum link model masquerading as deconfined criticality},
\emph{J. Stat. Mech.} (2013) P12010 

\bibitem{Evertz}
H. G. Evertz, M. Hasenbusch, M. Marcu, K. Pinn, and S. Solomon,
\emph{Cluster algorithms for surfaces},
\emph{Int. J. Mod. Phys. C} {\bf 03} (1992) 01

\end{thebibliography}
\end{document}